\documentclass[conference]{IEEEtran}
\ifCLASSINFOpdf
   \usepackage[pdftex]{graphicx}
   \graphicspath{{./pdf/}{./jpeg/}}
   \DeclareGraphicsExtensions{.pdf,.jpeg,.png}
\else
   \usepackage[dvips]{graphicx}
   \graphicspath{{./eps/}}
   \DeclareGraphicsExtensions{.eps}
\fi
\usepackage{algorithm}

\usepackage{algorithm,algpseudocode}
\usepackage{array}
\usepackage{float}
\usepackage{lipsum}
\usepackage{authblk}
\usepackage{authblk}
\usepackage[bookmarks=false]{hyperref}

\begin{document}
%
\title{Fingerprinting OpenFlow controllers: The first step to attack an SDN control plane}
%
%
%

\author[1]{Abdelhadi Azzouni\thanks{abdelhadi.azzouni@lip6.fr}}
\author[2]{Othmen Braham\thanks{othmen.braham@virtuor.fr}}
\author[1]{Nguyen Thi Mai Trang\thanks{thi-mai-trang.Nguyen@lip6.fr}}
\author[1]{Guy Pujolle\thanks{guy.pujolle@lip6.fr}}
\author[3]{Raouf Boutaba\thanks{E.E@university.edu}}
\affil[1]{LIP6 / UPMC; Paris, France  \{abdelhadi.azzouni,thi-mai-trang.nguyen,guy.pujolle\}@lip6.fr}
\affil[2]{Virtuor; Paris, France  othmen.braham@virtuor.fr}
\affil[3]{University of Waterloo; Waterloo, ON, Canada  rboutaba@uwaterloo.ca }
\maketitle

\begin{abstract}

Software-Defined Networking (SDN) controllers are considered as Network Operating Systems (NOSs) and often 
viewed as a single point of failure. Detecting which SDN controller is managing a target network is a big step for an attacker 
to launch specific/effective attacks against it.
In this paper, we demonstrate the feasibility of fingerpirinting SDN controllers.
We propose techniques allowing an attacker placed in the data plane, 
which is supposed to be physically separate from the control plane, 
to detect which controller is managing the network.
To the best of our knowledge, this is the first work on fingerprinting SDN controllers, 
with as primary goal to emphasize the necessity to highly secure
the controller. We focus on OpenFlow-based SDN networks since 
OpenFlow is currently the most deployed SDN technology by 
hardware and software vendors.


\end{abstract}
%
%

 {\bf { \it keywords - }}
Software-Defined Networking, OpenFlow, Control Plane, security.

%
\IEEEpeerreviewmaketitle

\section{Introduction and Motivation}


Software-Defined Networking (SDN) is an emerging architecture that is dynamic, agile, centrally managed and programmable, 
making it ideal for the evolving nature of today's applications. 
SDN separates the network control plane from the data plane (Fig. \ref{archi}), 
enabling more flexibility in managing and programming the network. 
The centralized control provided by SDN is expected to facilitate the deployment and hardening of network security \cite{survey1, survey2}.
However, SDN controllers can be subject to new threats
compared to conventional network architectures. 
For example, an attacker can change the whole underpinning of the network traffic behavior by modifying the controller. 
The Open Networking Foundation (ONF) identifies a number of SDN security issues that the community must address \cite{ONF-sec}:

\begin{figure}[h]
\centering
   \includegraphics[scale=0.2]{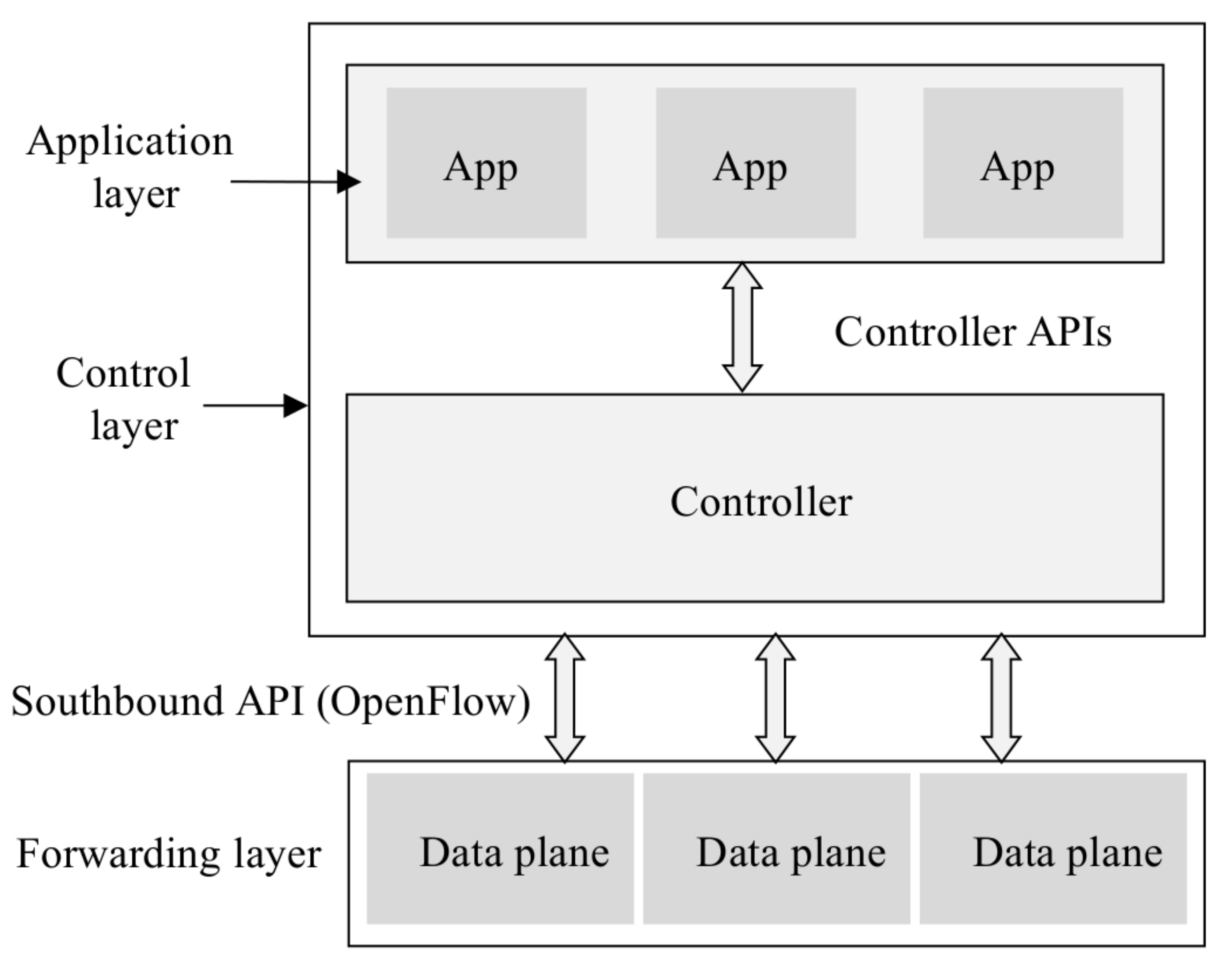}
   \vspace{-1em}
   \caption{\label{archi} SDN architecture}
   \vspace{-1em}
\end{figure}


\begin{itemize}
\item The centralized controller emerges as a potential single point of attack that must be protected.
\item The southbound interface between the controller and underlying networking 
devices (OpenFlow) is vulnerable to threats that could degrade the 
availability, performance, and integrity of the network. Using TLS or UDP/DTLS is recommended to secure the OpenFlow channel.
\item The underlying network must be capable of enduring occasional 
periods where the SDN controller is unavailable.
\end{itemize}

In the most common schemes of attacking a remote system, the first step is to determine 
the set of possible attacks by collecting information about the target.
In this paper, we demonstrate some techniques that allow an attacker 
to fingerprint the OpenFlow controller of the network. 
Once the attacker knows which controller is used, 
he/she can launch tailored attacks exploiting its known vulnerabilities.
We study the common case where the attacker
is placed in the underlying network managed by the target SDN controller, 
and does not have access to either the controller or the control channel. 

This work aims to demonstrate the feasibility of fingerpirinting OpenFlow controllers, 
with the ultimate goal of building a Penetration Testing framework 
that can be used by network administrators to test their SDN networks. Many frameworks have been created for the same purpose in traditional networks, NMAP \cite{nmap}, for instance,
is a widely used scanner that can fingerprint remote systems among other capabilities.
OWASP Zed Attack Proxy Project (ZAP) \cite{zap} is another security testing framework that includes fingerprinting remote web servers and web applications.
As in NMAP and ZAP, our proposed techniques are not to be used separately, 
that is, one may get non-accurate results when only using the first Timing-Analysis based technique (section \ref{toutinference}) for example,
but the combination of all proposed techniques, generally gives accurate results. 
The contributions of this paper are as follows:
\begin{itemize}
 \item We demonstrate the feasibility of fingerpirinting attack on OpenFlow controllers by designing, 
 implementing and testing several fingerpirinting techniques and
 \item We highlight the need for building a Penetration Testing framework for SDN networks.
\end{itemize}

This paper is organized as follows. Section \ref{section2} provides OpenFlow background information. 
Related works are discussed in section \ref{section3}. Our proposed fingerpirinting techniques are presented in section \ref{section4}, 
our experimental testbed is described in section \ref{section5}
and the results are given in section \ref{section6}. Section \ref{section7} concludes the paper and discusses some future directions.

\section{Background Information} \label{section2} 

OpenFlow is the first standard communication interface defined between the control and forwarding layers of an SDN architecture \cite{ONF-OF}.
Contrary to traditional routers, where the fast packet forwarding (data path) and the high level routing decisions (control path) occur on the same device. 
OpenFlow separates these two functions: An OpenFlow switch consists of one or more flow tables, which performs packet lookups 
and forwarding, and interfaces to external controllers. The controller controls the switch by
adding, updating, and deleting flow entries via OpenFlow messages, proactively or reactively (in response to arrival of new flows) .

An OpenFlow message is either a switch-to-controller or a controller-to-switch message. OpenFlow messages are detailed in \cite{OF-specif} 
of which the most important ones are:

\begin{itemize}
 \item Hello messages: exchanged between the controller and the switch when the connection is first established.
 \item Echo request/reply messages: used to exchange information about latency, bandwidth and liveness. 
 \item Packet-In messages: used by the switch to send a packet to the controller when it has no flow-table matching the packet.
 \item Packet-Out messages: used by the controller to inject packets into the data plane of a particular switch.
 \item Flow-mod messages: used by the controller to modify the state of an OpenFlow switch.
 \item Stats request messages: used by the controller to request information about individual flows.
\end{itemize}
%
%

\section{Related Work} \label{section3}
S. Shi and G. Gun developed SDN Scanner \cite{sdnscanner} which exploits the network header field change. 
If a client sends packets to an SDN network, this client will observe different response times, 
because the flow setup time can be added in the case of non-matching flow (i.e., there is 
no corresponding flow rule in the data plane: response time $T1$)  
compared to the case when the corresponding flow rule exists 
(response time $T2$). 
SDN Scanner collects the response times then uses statistical tests to compare them. 
Thus, if an attacker can clearly differentiate $T1$ from $T2$ then he/she can detect the SDN network (the presence of an SDN controller).
The evaluations conducted in the paper showed that SDN Scanner can fingerprint 24 networks out of 28 (i.e., a fingerprinting rate of 85.7\%). 
However, SDN Scanner does not detect the controller type. In addition, collecting accurate values of $T1$ and $T2$ is extremely hard in real-world WANs 
because of the many variables that affect the response time. As such this method may not be efficient in WANs. 
\cite{sdnfingerprinting2015} leverages information from  the  $RTT$  and  packet-pair  dispersion to fingerprint 
controller-switch interactions (i.e. whether  an  interaction
between the controller and the switches has been triggered by a given packet) in a remote SDN network.
%

L. Junyuan et. al \cite{inference} propose techniques to infer key network parameters like flow table capacity and flow table usage.
For example, when the flow table is full, extra interactions between controller and switch are needed to remove some of the existing 
flow entries to make room for new ones, which may result in a performance decrease of the network.
An attacker can take advantage of the perceived performance change to launch more effective attacks. More 
specifically, knowing the flow table size and usage, the attacker can estimate with high accuracy how many packets
he/she needs to generate per second to flood the flow table and the required time to fill it up. 
hence, he/she could choose and correctly configure their attacking tools. Contrary to
\cite{inference}, our methods aim to infer 
control plane parameters to fingerprint controllers, which is more critical and of higher impact. 

\section{Fingerprinting OpenFlow Controllers}\label{section4}
The main approach developed in this paper is to combine several techniques to fingerprint an SDN controller 
from its underlying data forwarding plane.
Although our proposed techniques can be used separately, the accuracy of the results is much higher 
when combining them. Also, using only one technique may not give any result in some situations. 
In other words, each method has its success probability, 
and combining several techniques intuitively increases the probability of identifying the type of SDN controller used. 

The following subsections present our techniques categorized into two classes: 
Timing-Analysis based techniques and Packet-Analysis based techniques.
\subsection{Timing-Analysis based techniques}
These techniques are based on time measurement to infer some indicative parameters of the controller. 

\subsubsection{Timeout Values Inference}\label{toutinference}
    
Each flow entry has an idle\_timeout and a hard\_timeout field values associated with it.
They indicate respectively the time in seconds after which the entry will be removed 
from the switch if no packet matches it, and the time after which 
to remove the entry anyway. These timeout values can be set and modified by application developers or 
network administrators. But, in most cases when the network or parts of the network only need a basic flow forwarding without additional 
traffic engineering logic, the network admins tend to use the forwarding applications that come with the controllers (typically
L2-Switches) and the probability that they change these applications' parameters is fairly low.
Note that in recent controllers, those forwarding elements even include some advanced features \cite{floodl2}.

The idea is to infer flow-entry timeout values and compare them to known timeout values of different controllers (timeout database).
The timeout database is constructed as follows: for open source controllers, default timeout values can be gathered from their code source or configuration files.
For proprietary controllers, the default timeout values can easily be figured out by simply using the controller and directly measuring the values.
This method can be fairly accurate because of the low probability for default values to be modified by administrators.

To measure timeout values from an end-host in the underlying network, 
we propose the two following algorithms (algorithm \ref{fingerpirinting.alg1} 
and algorithm \ref{fingerpirinting.alg2}). These algorithms consider network disruptions that may affect 
communication channels between end-hosts and the switch, and between the switch and the controller.  
Both algorithms require the ability to connect to another end-host in the same data plane (a pingable end-host). 
Algorithm \ref{fingerpirinting.alg1} measures idle\_timeout in two steps: first,
it calculates $RTT\_{avg}$ (average Round-Trip Time using ping) in case when corresponding flow entries are installed in the switch. 
Measurements may be made for a configurable duration and/or number of probing packets $n$. Second, 
it measures $RTT$ every $wait$ seconds. $wait$ value will be incremented by $step$ seconds until a significant difference between
measured $RTT$ and calculated $RTT\_{avg}$ is encountered. This difference means that the flow entry expired and the switch needed to call 
the controller asking how to handle the new ping. Final value of $wait$ matches the flow-entry $idle\_timeout$ value.
A more accurate version of the algorithm is conceivable by using a binary search around the final $wait$ value, 
but by using $step$ of $5 ms$, the algorithm remains very accurate even without binary search.

Note that in some controllers, the default $idle\_timeout$ value is set to $0$ which means 
infinite, so the flow entry will never be removed. We found this in the $Ryu$ controller and $Hydrogen$, an old version of $OpenDaylight$ \cite{odl}.
In this case, after a number of iterations, the algorithm will decide that the $idle\_timeout$ value is infinite and the controller may
be $Ryu$ or $Hydrogen$ version of $OpenDaylight$. The search space has been limited to two controllers in this case, 
but we need to apply more techniques to decide which one of them. 

\begin{algorithm}[H]
\caption{idle\_timeout measurement}
\label{fingerpirinting.alg1}
\begin{algorithmic} [1]
  \State Send first ping to install flow entry;
  \State Send $n$ pings and calculate the average ping time $RTT\_{avg}$;
  \State Wait $wait$ seconds;
  \State Send one ping and calculate ping time $T_{ping}$
  \If  {$T_{ping} \approx RTT\_{avg}$ } //the flow entry still exists
    \State $wait \leftarrow wait + step$;
    \State Go to 3;
  \Else   //idle\_timeout expired and the flow entry removed
    \State  $idle\_timeout = wait$
  \EndIf
\end{algorithmic}
\end{algorithm}

To measure $hard\_timeout$ value, we first calculate the average of $RTT$ time $(RTT\_{avg})$ 
and $idle\_timeout$ values as in algorithm \ref{fingerpirinting.alg1}. Second, we send one ping to
install the flow entry in the switch. Then, we send a ping every $wait$ seconds such as $wait$ value 
is less then $idle\_timeout$. As long as the $RTT$ value is close to the average $(RTT\_{avg})$, we continue to add
$wait$ seconds to the $hard\_timeout$ value initialized to zero. We stop when 
we find a $RTT$ value which is significantly greater than $(RTT\_{avg})$.

\begin{algorithm}[H]
\caption{hard\_timeout calculation}
\label{fingerpirinting.alg2}
\begin{algorithmic} [1]
  \State $hard\_timeout$ $\leftarrow$ 0 seconds;
  \State Calculate $RTT\_{avg}$ as in algorithm 1;
  \State Calculate $idle\_timeout$ as in algorithm 1;
  \State Send one ping to make the controller install flow entry;

  \State Wait $wait$ seconds, $wait$ must be less then $idle\_timeout$;
  \State Send one ping and calculate ping time $T_{ping}$;
  \If  {$T_{ping} \approx RTT\_{avg}$ } //the flow entry still exists
    \State $hard\_timeout \leftarrow hard\_timeout + wait$
    \State Go to 5;
  \Else   //hard\_timeout expired and the flow entry removed
    \State  print $hard\_timeout$
  \EndIf
\end{algorithmic}
\end{algorithm}

The attacker then compares the measured values $(idle\_timeout, hard\_timeout)$
to known timeout values of controllers to guess which controller is used. \\

\subsubsection{Processing-Time Inference}
Each SDN controller is programmed differently using different tools, libraries and frameworks, 
so that each controller has its own execution speed. In other words, when receiving packets from the data plane,
each controller takes a different time to process those packets and reply back to the data plane. 
The idea of this technique is to use estimated 
packet-processing time to determine the controller.
As we mentioned before, 
authors of \cite{sdnscanner} used timing to determine if a remote network 
is an SDN network based on the difference of $RTT$ in two cases: presence 
and absence of flow entries.
As it has been mentioned by the authors, it is very difficult to measure with high accuracy the $RTT$ to
a remote network in a WAN because of many potential sources of disruption that may
result in random variations of $RTT$ values. In our technique, these disruption sources are minimal since the attacker
is placed in the data plane of the target controller. And unlike \cite{sdnscanner}, our method uses some
key parameters inferred from the network to estimate processing time with higher precision.

The main idea in our approach is to measure the
response time of the target controller and compare it to the processing-time database created beforehand. 
The processing-time database is a table that 
associates each controller to its processing time. 
Like in the previous technique (timeout values inference), we need a pingable destination end-host 
in the same data plane (the best scenario is that the attacker controls the destination end-host as well to be sure that its processing 
time does not affect the measurements).
\begin{figure} [h]  \setlength{\textfloatsep}{5pt}
\centering
   \includegraphics[scale=0.2]{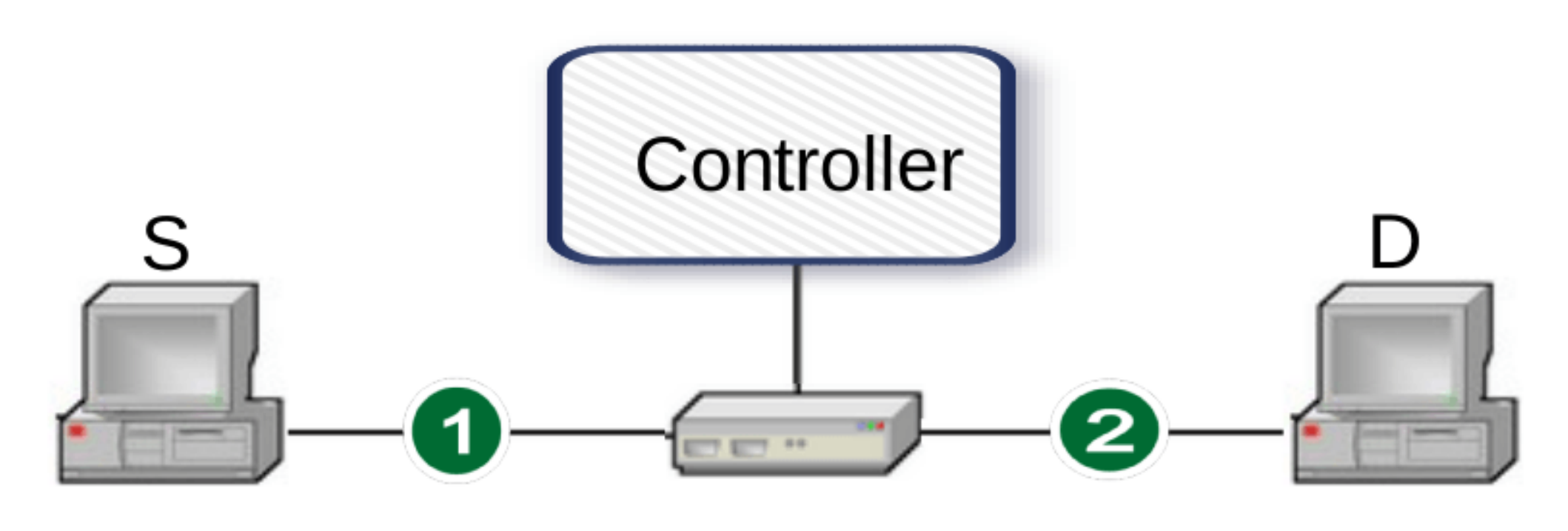}
   \vspace{-1em}
   \caption{\label{fig:simplearch} Simplified architecture to measure controllers' processing time}
   \vspace{-1em}
\end{figure} 
To create the processing-time database, we use a simplified architecture (Fig. \ref{fig:simplearch}) where the propagation times $(1)$ and $(2)$ are minimal.
we first measure $idle\_timeout$ and $RTT\_{avg}$ values
as in algorithm \ref{fingerpirinting.alg1}. Then, we send $n$ (100 for example) pings separated by $period$ seconds
between every two pings, with $period$ greater then $idle\_timeout$. Every ping will cause
the switch to send a Packet-In to the controller (by receiving the Packet-In message,
the controller processes it to extract field values and installs the corresponding flow rule into 
the switch). Finally we calculate the average ping time $T_{pavg}$ of the $n$ pings and we record 
$T_{pavg}-RTT\_{avg}$ value 
in a table. This is the processing time of the current controller. 
We repeat this process with all controllers and we create the processing-time database
by inserting tuples $(controller, processing\_time(T_p))$ (algorithm \ref{fingerpirinting.alg3}).

\begin{algorithm}[H]
\caption{Building the processing-time database}
\label{fingerpirinting.alg3}
\begin{algorithmic} [1]
  \State Calculate $RTT\_{avg}$ as in algorithm 1;
  \State Calculate $idle\_timeout$ as in algorithm 1;  
  \For {$i \leftarrow 1 .. n$}  
    \State Wait $period$ seconds, $period$ must be greater than $idle\_timeout$;
    \State Send a ping and save ping time;
  \EndFor
  \State Calculate the average of saved ping time values $T_{pavg}$ and calculate controller processing time 
  $T_p = T_{pavg}-RTT\_{avg}$; 
  \State Insert $(controller, T_p)$ in the processing-time database;
\end{algorithmic}
\end{algorithm}
Note that, as propagation times $(1)$ and $(2)$ (Fig. \ref{fig:simplearch}) are minimal, 
measured $RTT\_{avg}$ is accurate and hence $T_p$ values are accurate.

\begin{algorithm}[H]
\caption{Fingerprinting $controller$}
\label{fingerpirinting.alg4}
\begin{algorithmic} [1]
  \State Calculate $RTT\_{avg}$ as in algorithm 1;
  \State Calculate $idle\_timeout$ as in algorithm 1;  
  \For {$i \leftarrow 1 .. m$}  //m = 20 for example
    \State Wait $period$ seconds, $period$ must be greater then $idle\_timeout$;
    \State Send a ping and save ping time;
  \EndFor
  \State Calculate the average of saved ping-time values $RTT'$ and compare $RTT' - RTT\_{avg}$ to the processing-time entries;
\end{algorithmic}
\end{algorithm}

Now that we have the processing-time database, 
to fingerprint the target controller that manages the real SDN network we are connected in, we first measure $RTT\_{avg}$ 
to a destination, then we ping the same destination
with a spoofed IP address to ensure that no corresponding flow entry exists in the switch
and we compare the value $RTT - RTT\_{avg}$ to the processing-time database entries. 
For accuracy, we do not rely on a single ping,
disruptions can happen during the ping affecting the response time. 
Instead, we send many (20 for example) pings with $period$ seconds between every 
two pings (such as $period$ value is greater than $idle\_timeout$), we calculate the average of 
these ping times $RTT'$ and finally compare the value $RTT' - RTT\_{avg}$ 
to the processing-time database entries (algorithm \ref{fingerpirinting.alg4}).

In addition to the probability that the network admin somehow modifies the execution time of the controller, which we argue is very low,
there is a fair chance that during the scan, the controller is overloaded resolving requests and installing rules, 
which may significantly change the response time.
In this case, if the attacker has further knowledge about the network state then he/she can surpass this problem. 
For example, he/she can avoid peak hours, 
and only scan the controller when the network
is in its normal state. 


\subsection{Packet-Analysis based techniques}
\subsubsection{LLDP message analysis}

This is a passive method which consists of identifying the controller by sniffing and analyzing
OpenFlow Discovery Protocol $(OFDP)$ packets sent over the data plane.

SDN is based on maintaining a global network view at the level of the controller.
To obtain the global network topology, discovery modules of the controllers use $OFDP$ to collect updated 
information from different elements of the network including end hosts. 
$OFDP$ leverages the packet format of Link Layer Discovery Protocol $(LLDP)$
with subtle modifications to perform
topology discovery in an OpenFlow network. 

Unlike ordinary $LLDP$ enabled switches, an OpenFlow switch 
needs the controller to send and process $OFDP$ messages and cannot do this by itself.
The following is a simple scenario of the topology discovery process using $OFDP$.
First, the SDN controller creates an individual LLDP packet for each port on each switch.
Then, the controller sends these packets to the switches via Packet-Out messages that 
include instructions to send them out on the corresponding ports. 
In each switch, all received $LLDP$ packets will be forwarded to neighbours. 
When a switch receives a new $LLDP$ packet from another switch, 
it forwards it to the controller via a Packet-In message.
At the end of the process, the controller will get information about all the data-plane connections.
The entire discovery process is repeated periodically with the time periods varying from 
one controller to another, which is can be leveraged to identify which controller is managing the network.
Also, the content of the $LLDP$ packets differs from one controller to another, which can be used accurately identify the controller.   
Table \ref{fingerpirinting.tablelldp} in section \ref{results:lldp} shows $LLDP$ packets sent by different controllers.\\


\subsubsection{ARP response analysis}

This technique can only be used to determine 
if the controller is the Hydrogen version of OpenDaylight and cannot be generalized to other types of controllers. 
It builds on the observation
of how the controller reacts to unknown Address Resolution Protocol $(ARP)$ requests in the data plane.
The attacker sends an unknown $ARP$ request, which means that the destination IP address is 
not assigned to any host in the network.
As the destination IP is not present in the network, the switch, 
in addition to broadcasting the request, sends it to the SDN controller via a 
Packet-In message asking how to handle it. 
The OpenFlow specifications indicate that the controller responds to the switch by a 
Packet-Out and/or a flow-mod message explaining how to handle the request. 
The controller's response message differs from one controller to another, 
but the only controller whose behavior can be captured from an end-host is Hydrogen.
Hydrogen version of OpenDaylight
instructs the switch to broadcast the request once again which duplicates it in the broadcast domain. 
This duplicated ARP request, with one of the switch's Media Access Control $(MAC)$ addresses as source address, 
indicates that Hydrogen is used.

As we mentioned in the introduction, the techniques we presented in this section are not to be used in an exclusive manner. 
Each technique is able to identify the controller with a certain probability that we did not compute analytically in this paper. 
The user can use a subset or all the techniques executing them one by one, or better combine them in some optimal order.
The selection of the optimal combination of techniques is an interesting research question that we leave for future work.

\section{Experiment Environment and Methodology}\label{section5}

\begin{figure} [h]
\centering
   \includegraphics[scale=0.2]{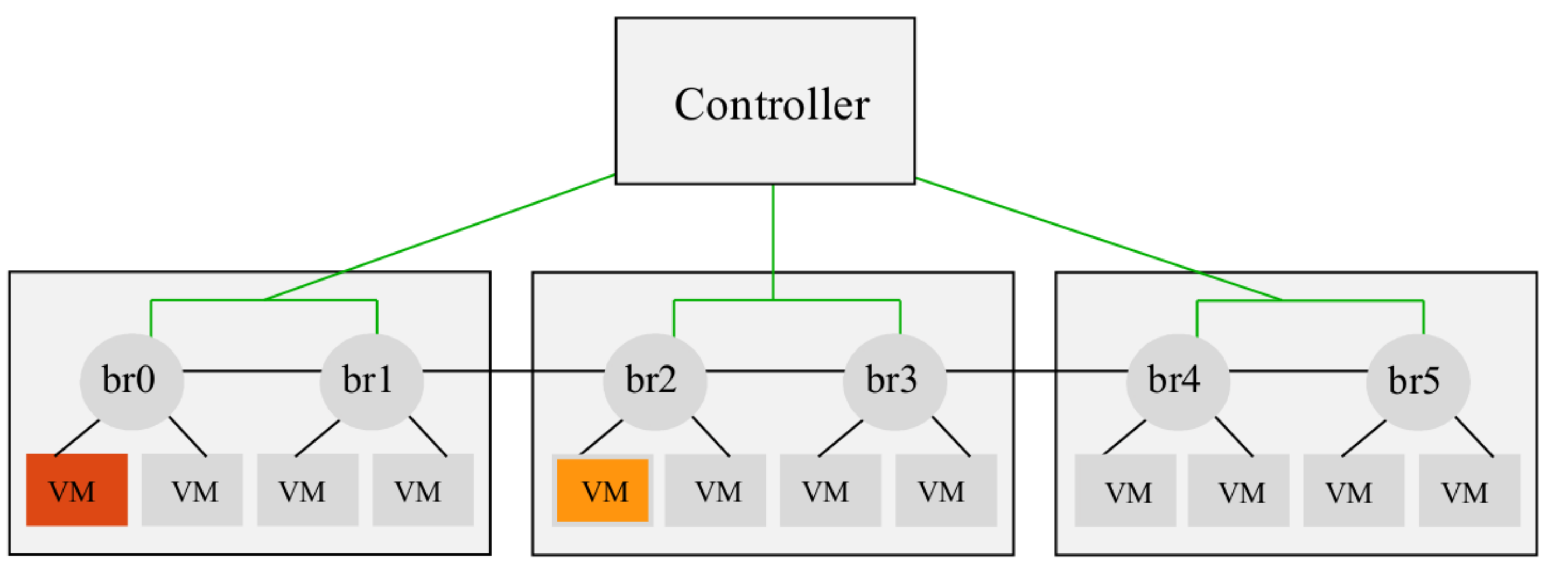}
      \vspace{-1em}
   \caption{\label{fig:test-env} Test environment}
   \vspace{-1em}
\end{figure}

As shown in Figure \ref{fig:test-env}, our experiment environment consists of four physical 
machines (only three are shown in Fig. \ref{fig:test-env}) carrying 4 virtual machines each and connected via OpenFlow virtual bridges
(Openvswitch) forming a small-size data-center where VMs generate random traffic (ping and iperf) to random destinations.
Note that, since we are not exploiting any weaknesses in the switch, it does not make any difference using a virtual switch or a physical one in this context, 
we only need a switch that correctly implements OpenFlow specifications.
Note also that we did not add hops (transit switches) between bridges and the controller because even in real-world networks, 
a very small number (0, 1 or 2) of transit switches is enough to build a fairly large Local Area SDN network, like a data-center or a campus network. 
Such a small number of hops does not affect timing measurements in previous algorithms, 
The attacker is on the red (or black) VM connected to $br0$, and can ping the orange (or dark grey) VM connected to $br2$ (it could be any other VM in the network).

We have performed our experiments on five open source still maintained, OpenFlow controllers among the most widely used:
OpenDaylight \cite{odl},
POX \cite{nox},
Beacon \cite{beacon},
Floodlight\cite{floodlight},
and Ryu \cite{ryu}.

\section{Results}\label{section6}

\subsection{Timeout Values Inference technique}
To evaluate the Timeout Values Inference method, 
we ran algorithms \ref{fingerpirinting.alg1} and \ref{fingerpirinting.alg2} ten times on each controller
from the set of our target controllers. Default timeout values are given in table \ref{tableidle}. Our algorithms
did 2 errors in $50$ measurements in both $idle\_timeout$ and $hard\_timeout$ and that is because
algorithm \ref{fingerpirinting.alg1} is used in algorithm \ref{fingerpirinting.alg2}.

\begin{table}[h]
\begin{center}
\begin{tabular}{|p{2cm}|p{2cm}|p{3cm}|p{3cm}|p{4cm}|}    
    Controller & 
    idle\_timeout (s)&
    hard\_timeout (s)\\
    \hline
    OpenDaylight &
    0 &
    0 \\
    \hline
    Floodlight &
    5 &
    0 \\
    \hline
    POX &
    10 &
    30 \\
    \hline
    Ryu &
    0 &
    0 \\
    \hline
    Beacon& 
    5 &
    0
\end{tabular}
\caption{\label{tableidle}Default Timeout Values}
\vspace{-1em}
\end{center}
\end{table}

We also evaluated algorithms \ref{fingerpirinting.alg1} and \ref{fingerpirinting.alg2} separately
by manually setting different values for $idle\_timeout$ and $hard\_timeout$ in POX source code, and
running the algorithms from the attacker virtual machine (red VM in Fig. \ref{fig:test-env}) to infer these values.
We set the values $5, 10, 15, .. 30 ms$ for $idle\_timeout$ and the values $10, 20, .. 60 ms$ for $hard\_timeout$ respectively. 
For each algorithm, we repeated the execution $10$ times on each value.
$idle\_timeout$ calculation algorithm has an error 
rate of $0.03\%$ (2 errors in $60$ measurements) with a relative error of less than $1s$.
The $hard\_timeout$ calculation 
algorithm has an error rate of $0\%$ (no error) on $60$ measurements.
%


\subsection{Processing-Time Inference technique}
First, we have built the processing-time database (table \ref{fingerpirinting.tabletp}) of our set of target controllers 
by running algorithm \ref{fingerpirinting.alg3} ($n = 100$) on a simplified testbed as described in Fig. \ref{fig:simplearch}. 

\begin{figure} [h]
\centering
   \includegraphics[scale=0.25]{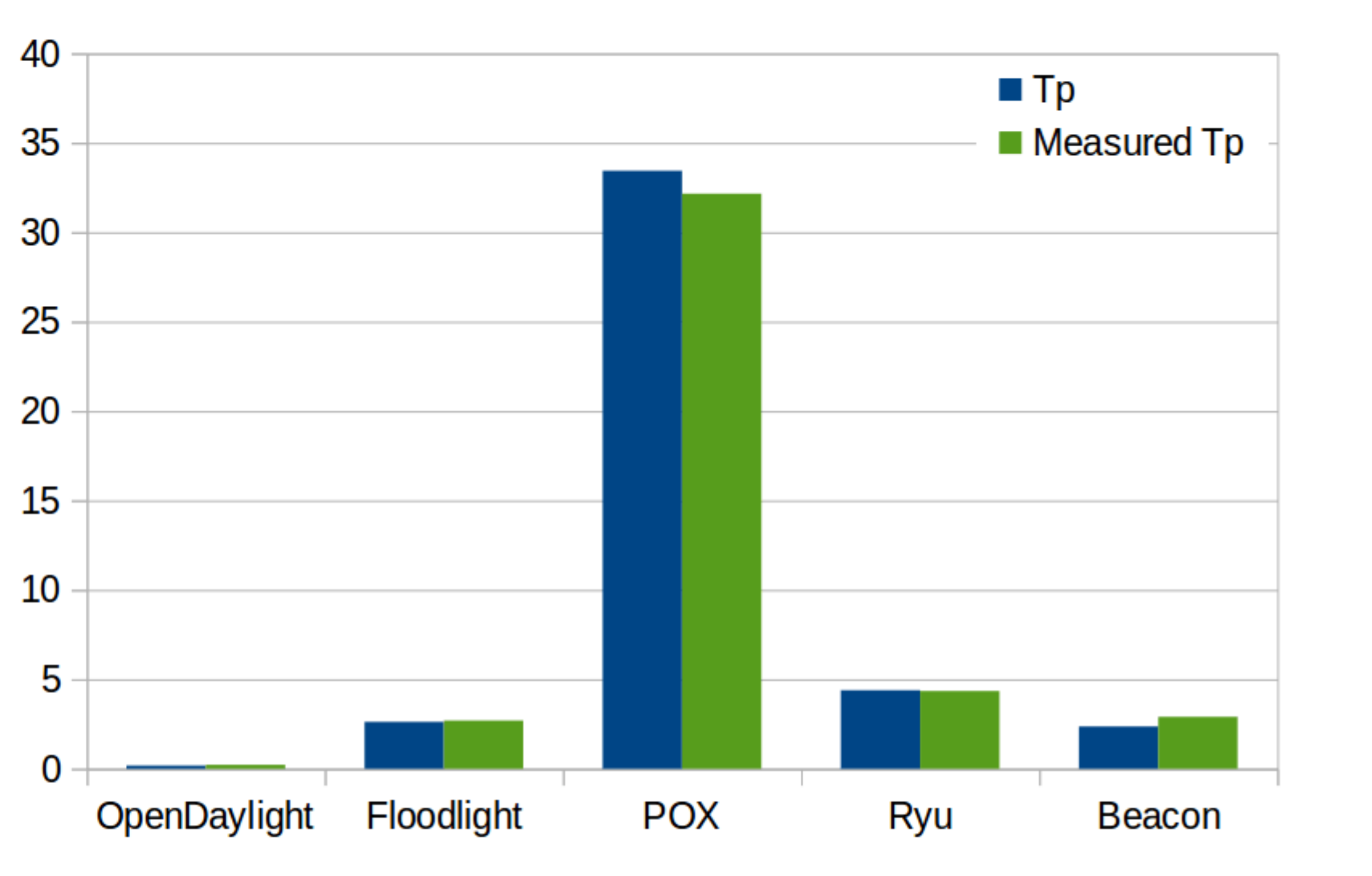}
   \caption{\label{fig:resulttp} Measured processing times compared to average processing times}
   
\end{figure}

\begin{table}[h]

\begin{center}
\begin{tabular}{|p{2cm}|p{2cm}|p{3cm}|p{3cm}|p{4cm}|}    
    Controller & 
    $T_p$ $(ms)$&
    $T_p$ adjusted $(ms)$\\
    \hline
    OpenDaylight &
    1.004 &
    0.177 \\
    \hline
    Floodlight &
    3.454 &
    2.627 \\
    \hline
    POX &
    34.266 &
    33.439 \\
    \hline
    Ryu &
    5.216 &
    4.389 \\
    \hline
    Beacon& 
    3.197 &
    2.370
\end{tabular}
\caption{\label{fingerpirinting.tabletp}Processing-time database ($T_p$: processing time).}
\vspace{-1em}
\end{center}
\end{table}
%
%
Then, to evaluate this technique in our experimental environment (Fig. \ref{fig:test-env}) we have run algorithm \ref{fingerpirinting.alg4} ten times: 
measured $T_p$ in Fig. \ref{fig:resulttp} is the average value of the different executions.
To get more precise comparisons, we calculate 
$T_p$ adjusted: adjusted processing time = processing time - the average of $RTT$ time in case the flow rule exists ($RTT_{avg}$).

For controllers Floodlight and Beacon which
have very similar values of $T_p$, the use of only this technique, is not sufficient as it cannot decide between them. 

%
%

\subsection{$LLDP$ message analysis technique} \label{results:lldp}
\begin{figure} [h]
\centering
   \includegraphics[scale=0.3]{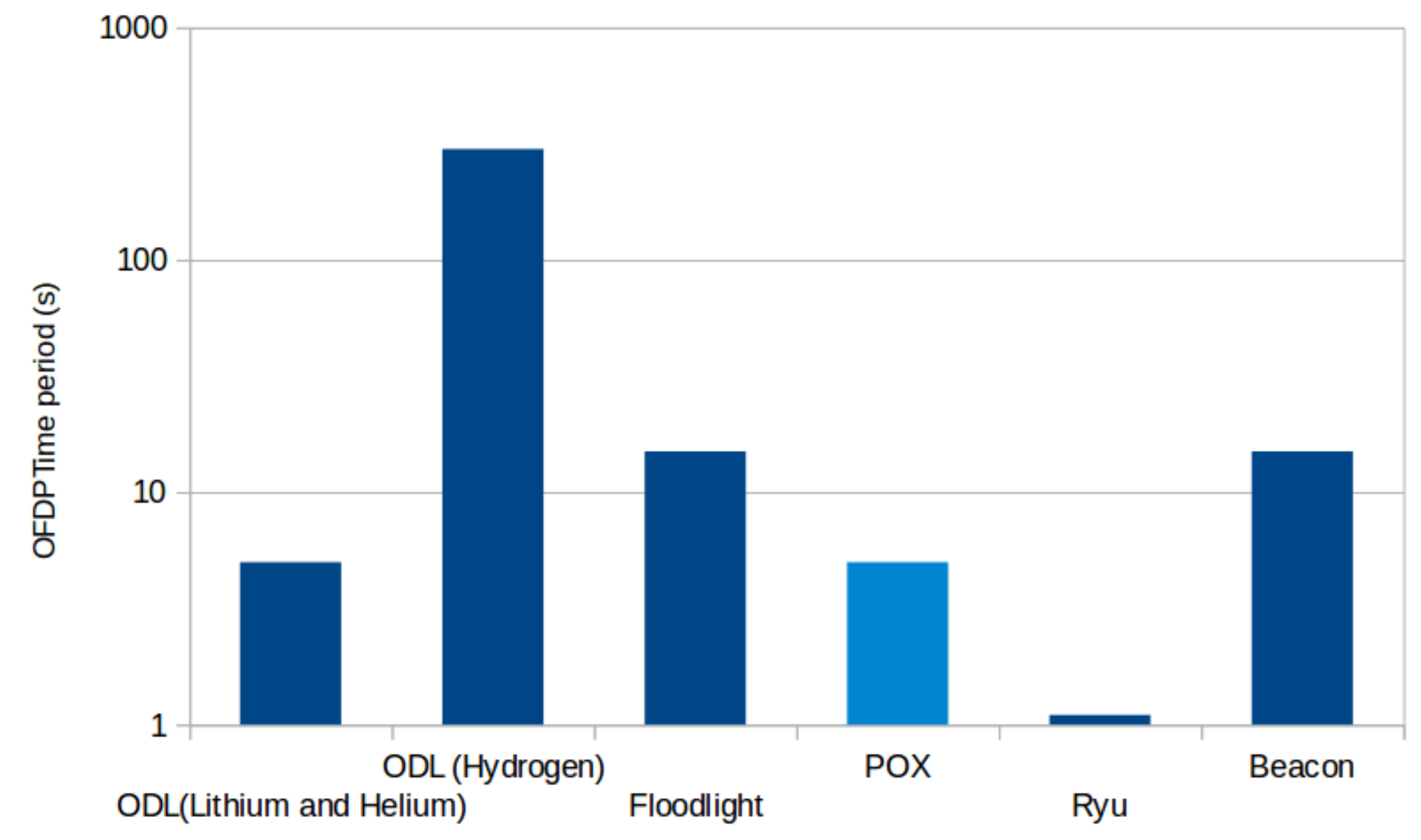}
   \caption{\label{fig:lldp} Controllers' LLDP-emission-interval comparison}
\end{figure}

Figure \ref{fig:lldp} compares $LLDP$-packet reception intervals for different controllers.  
Table \ref{fingerpirinting.tablelldp} shows the difference between controllers' $LLDP$ packets.
By receiving the $LLDP$ packet, the attacker compares the different values against 
Fig. \ref{fig:lldp} and table \ref{fingerpirinting.tablelldp}
to identify the controller. Similar to technique \ref{toutinference}, 
for proprietary controllers, the way to gather $LLDP$ information is to 
simply use the controllers, analyze its $LLDP$ packets, then use this information to fingerprint target controllers.

\begin{table}
\begin{center} 

\begin{tabular}{|p{2.5cm}|p{2.5cm}|p{10cm}|p{3cm}|p{4cm}|}
    Controller & 
    $OFDP$ interval $(s)$&
    Remarks  \\
    \hline
    OpenDaylight (Lithium \& Helium) &
    5 &
    LLDP packets include System Name field with value = "openflow" and no System Description field \\
    \hline
    OpenDaylight (Hydrogen) &
    300 &
    LLDP packets include System Name field with value = "OF|[MAC address of the OF switch]" and no System Description field \\
     \hline
    Floodlight& 
    15 &
    Each LLDP packet is followed by an 0x8942 Ethernet packet sent in broadcast. 
    This makes it easy to distinguish between Floodlight and Beacon \\
    \hline
    POX &
    variable ($\approx$ 5) &
    LLDP packets include System Description field with value = "dpid:[MAC address of the OF switch]"\\
    \hline
    Ryu &
    1 &
    Note that the Topology discovery module is still not stable and not included in the controller core.
     \\
    \hline
    Beacon& 
    15 &
    LLDP packets include two "unknown" fields and no System Name or Description feild\\
    
\end{tabular}
   
\caption{\label{fingerpirinting.tablelldp}Results of $LLDP$ message analysis}

\end{center}
\end{table}

\section{Conclusion}\label{section7}
In this work, we demonstrated the feasibility of fingerprinting attacks on OpenFlow controllers from the data plane
by designing, implementing and testing practical techniques to identify the controller without access to the control plane.
%
%
This is a critical step for a number of attack models since it provides the attacker with sufficient 
information about the controller to carry out more tailored attacks.
Knowing the vulnerabilities of the target controller or one of its components,
the attacker can indeed use known or design new attacks to take down the controller.
In the future, we plan to expand the scope of this work by fingerprinting a larger set of controllers and by designing more techniques 
for fingerprinting controllers. We also plan to investigate formal methods for the evaluation of fingerprinting techniques and 
how they can be possibly combined to increase success rate. Finally, we plan to explore what countermeasures must be  deployed 
to harden the security of SDN networks against controller fingerprinting and subsequent attacks. \\ \\ \\ \\  \\ \\ \\ \\ \\  \\ \\ \\ 

\end{document}